\documentstyle[11pt,aasms4]{article}
\def\etal{{\it et al.\ }}
\def\eg{{\it e.g.,}}

\begin{document}
\title{Saving Astronomical Treasures}

\author{Noah  Brosch{\footnote{Email: noah@wise.tau.ac.il}} }
\affil{The Wise Observatory and 
the School of Physics and Astronomy,
Raymond and Beverly Sackler Faculty of Exact Sciences,
Tel Aviv University, Tel Aviv 69978, Israel}

\author{Ren\'{e} Hudec{\footnote{Email: rhudec@asu.cas.cz}} }
\affil{Astronomical Institute, CZ-251 65 Ondrejov, Czech Republic}

\author{Peter Kroll{\footnote{Email: pk@stw.tu-ilmenau.de}} }
\affil{Sonneberg Observatory, Sternwarterstr. 32, D-96515 Sonneberg, Germany}

\author{Milcho Tsvetkov{\footnote{Email: mtsvetkov@hotmail.com}} }
\affil{Institute of Astronomy, Bulgarian Academy of Sciences, 72 Tsarigradskoe Shosee Blvd.,
BG-1784 Sofia, Bulgaria}

\begin{abstract}

The time domain is under-exploited in an astrophysical context,
because the large photographic plate collections are almost never
consulted by young astronomers. This is a result of lack of education 
of the more recent generations
of astronomers, and of the deluge of terabyte-size datasets produced
by new observations. Yet astronomical plates represent a unique set of resources
and their quality does {\bf not} improve with time.
We point out valuable scientific projects that
can be conducted with archival images and describe a first step to
realize this at the Uccle Observatory, in Belgium.

\end{abstract}

\section{Introduction}
Modern astronomy has become a multi-frequency discipline, where simultaneous observations
from the $\gamma$-rays to radio are used to disentangle the physics of celestial
sources. Such observations are obtained by ground or space-based platforms and
are available to the investigator almost immediately, in a digital format. Modern astronomy
has become a science domain where investigations are conducted by manipulating large
data files on computers.

Clearly, this was not always the case. Permanent records have always been kept by astronomers, from
scratches on pre-historic animal bones indicating the phases of the Moon, through stellar
catalogs of Greek and Roman astronomers (culminating with Ptolemy's {\it Megale Syntaxis} 
compilation)  and Galileo's
telescope drawings. Since more than a century ago, semi-permanent impartial records
have been available in the form of photographic plates. The first photograph of a comet 
(Donati in 1858) is almost 1.5 centuries old (Pasachoff \& Olson 1995).
Most of the images of the sky more than one
decade old are not available digitally, with the very few exceptions of the large sky surveys.
The scanning and subsequent calibration of the Palomar Sky Surveys (POSS-I and POSS-II)
yielded very useful data sets, such as the Guide Star Catalog and the
Digitized Sky Survey (DSS and XDSS). These digital
images can now be compared with images taken in other spectral bands.

Modern astronomy is, therefore, multi-spectral in that it combines optical information
with that collected in other bands. But while the recently-obtained data are available electronically
and can be used to explore physical processes currently taking place in the
target objects or recent changes in such objects, any long-term behavior remains largely 
unexplored.

Yet the world astronomical archives contain today more than two million photographic plates, each 
covering a sky region of one square degree or wider. Most of the plates are relegated to hardly-used
astronomical archives; some even represent obstacles to the development of existing observatories,
or belong to archives of ``defunct'' observatories. The purpose of this paper is to
emphasize the scientific potential of such a rich heritage, to give some examples of
recent usage and of possibly interesting future usage, to expound on ways to provide the information to
the astronomical community, and to explain about the first steps taken in this direction 
by any interested group.

\section{Plate archives and the IAU}

The IAU Working Group on Sky Surveys (WGSS) has been aware of the lack of attention 
generally paid by the 
 astronomical community to the existence of neglected plate archives and of the need 
to safeguard their continued existence. These questions have been discussed at length during
meetings of the Organizing Committee of the WGSS and its previous incarnations. In particular,
this subject was emphasized by Dr. Richard West, who called on the WGSS to maintain a
living archive that could be used to identify and locate relevant plate material for
various research projects.

This project has now been realized by MT and his co-workers at the Astronomical Observatory
of the Bulgarian Academy of Sciences.
The information about many astronomical plate archives is contained in the Sofia Wide-Field
plate archive (http://www.skyarchive.org) and  has been described by Tsvetkov \etal (1997).

On the other hand, changes within old observatories are endangering many plate collections.
The lack of space caused the compression of the Harvard  (HCO) plate archives from three floors
in a building to two floors. In the process of moving, it is possible that some plates were 
broken and the information once stored in them is now lost (true, perhaps, of some 2\% 
of the plates in the collection of the HCO). 
On the closure of the Royal Greenwich Observatory in the UK, the
entire historical archive was transferred to temporary storage
in London. Although access to the collection, which is fully cataloged, is still available 
through the Institute of Astronomy in Cambridge, PPARC in the UK is seeking a long-term
home for the collection which will allow more research to be carried out and a program of plate digitization, where appropriate.

An IAU Working Group concerned with 100-yr old plates has been in existence for some years. Its
main target is the salvaging of the original plates of the Carte du Ciel (CdC), some of which are 
more than older than 100 years. Some of the CdC plates are already threatened because
 their emulsions are delaminating from the glass substrate, so their information is about to be lost forever to humanity. 
At the 24th General Assembly, the International Astronomical Union recognized the need to 
save the legacy of old photographic plates when it adopted Resolution B3, ``Saving 
the information in photographic observations''. The Resolution recognizes the 
need for judicious action, and urges the Union to support suitable steps
to save the plate collections. A Working Group is now being organized  (WG for Preservation 
and Digitizition of Photographic Plate Archives), which will include representatives of 
countries hosting research quality plates,
people engaged professionally in scanning and/or preserving plates, and leaders of 
projects to that end. This new WG is expected to operate under IAU Commission 5.

\section{Projects using archival information}

Below we give a number of examples when archival information from photographic plates was
instrumental, and in some instances essential, to disentangle astronomical information. We
also present a first-class scientific study that could be completed solely with archival
plates.

\subsection{Solar system studies}

{\bf Pluto:}
While it is well-known that the discovery of Neptune was the result of its perturbing 
effects on the orbit
of Uranus, the discovery of Pluto by Clyde Tombaugh was the result of a systematic search using
photographic plates. The orbit of Pluto is problematic because the observations span only
a small part of its orbit and the astrometry is prone to systematic zone errors in the catalogs.
In addition, there is a systematic bias whose origin is not known (Standish 1996). An
improvement in the absolute ephemeris is possible from remeasuring archival plates with new equipment
and putting the results in the framework of modern astrometric catalogs;
this is necessary for a future spacecraft mission to Pluto, such as the Pluto Fast Flyby Mission.

\subsection{Stellar studies}

{\bf R Aquari jet:}
The morphology of this object has been studied by Hollis \etal (1999) using archival information 
from the Lowell Observatory plate archives. The data from the digitized images, show that the 
jet of R Aqr was already
affecting the inner compact nebulosity even before the early 1970s, when it was first
observed and reported. The age of the jet can be estimated at $\sim$100 yr,  from
the proper motions of the oldest radio-jet components.

{\bf Eclipses in SS Lac:}
This system was known as an eclipsing one (P=14$^d$.42) until $\sim$1950. The
cessation of the eclipses was explained by Torres \& Stefanik (2000) as a result of a
third body orbiting in the system; this changed the inclination of the eclipsing binary. That
result was derived from studies of long-term photometric variability using photographic
plates as well as new observations.

{\bf The mass of Procyon:}
The derived mass for the F5 primary in the $\alpha$ CMi system has been adjusted  
by Girard \etal (2000), on the basis of astrometric information from more than 
250 archival plates. It is now close to the theoretical value.

{\bf Activity of red dwarf stars:}
Studies of variability of flaring stars can benefit very much from long-term monitoring with
photographic plates. Bondar' (1995) studied this with plates from Moscow (Sternberg), Odessa 
and Sonneberg observatories, covering a 90-yr baseline. Ten objects out of a sample of 29 
showed detected or suspected long-term variability. Two stars, V833 Mon (dK5e) and PZ Mon 
(dK2e),  showed high-amplitude
changes of 0.6 and 1.0 mag on timescales of 60 and 50-yr, respectively.

{\bf Searches for optical counterparts of GRBs:}
The existence of extensive archives on photographic plates makes it possible to search for
transient optical phenomena that could be related to recurrent activity of gamma-ray
bursts. Although such coincidences are likely to be extremely rare, Hudec \etal (1994) 
found such a transient by searching through 9104 photographic images covering the region of 
GRB 910219. They discovered that a short light flash ($<$35 minutes) occured on May 10, 
1905 and it could not have been
artificial. The recently suggested relation between GRBs and star forming regions in 
distant galaxies indicates that such recurrent events, from nearly the same sky location, are possible.

Previously, Shaeffer (1990) used the Harvard College Observatory plate archive to identify three optical
transients in error boxes of GRBs. A measure of confidence in the 
detections was obtained by performing similar searches for optical transients 
(with negative results) in a much larger
control sky area.

{\bf Optical analyses of X-ray stars:} 
For many of the optically identified X-ray stars, such as X-ray binaries, X-ray novae, cataclysmic
variables, etc., long-term optical data represent an extremely significant input for their
understanding. The Her X-1/HZ Her system may serve as such an example. Archival plates have revealed
the existence of active and inactive states, while the behavior of the system in an inactive 
state could be studied so far {\bf only} by using data from plates (Hudec \& Wenzel 1976, 
Simon, Hudec \& Kroll 2000)

\subsection{Photometric monitoring of AGNs}
It is well-known that many AGNs, and QSOs in particular, are optically variable. The value of
photographic plate archives for monitoring this form of acivity across historical time-scales 
has been
recognized for some time (\eg \, Angione \etal 1981). In fact, Angione \& Smith (1985)
used a subset of  historical photographs containing the 3C273 field to study its photometric
variability over a 93-yr period, from 1887 to 1980. A very interesting finding was the apparent
presence of a $\sim$16-yr periodicity in the light curve, as shown by a Fourier analysis of
the data from 1928 to 1954 (see Figure 1).

\begin{figure} 
\vspace{7cm}
\includegraphics{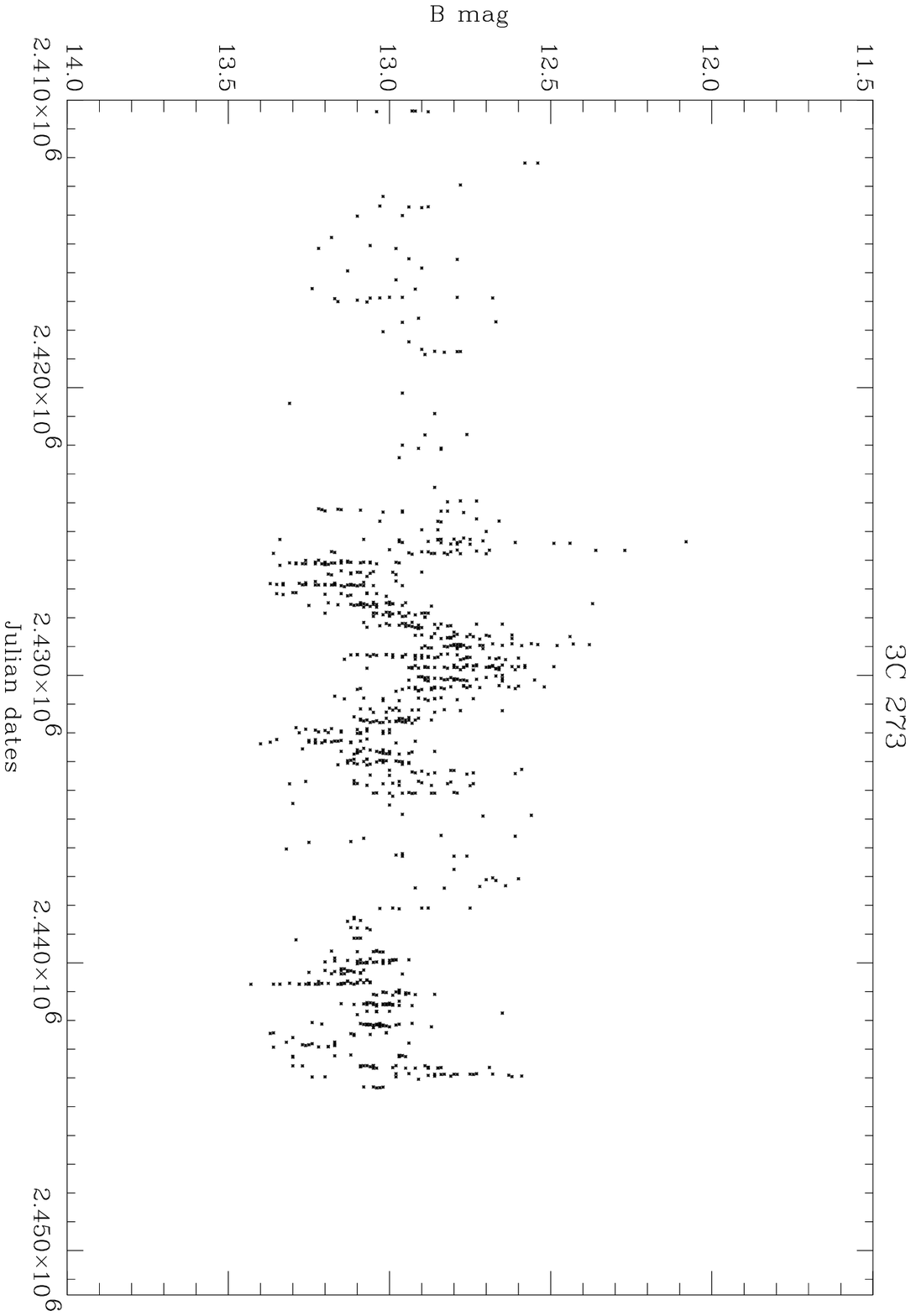}
\caption{Light curve for 3C273 using the data of Angione \& Smith (1985). Note the
two peaks, corresponding approximately to 1939-40 and 1949-52.}
\end{figure}

Angione \& Smith (1985) could not conclude whether that periodicity is real or not, nor 
whether it existed prior to 1928, owing to the sparseness of the data sets. Determining the 
reality of very long-term periodic variability is important for reverberation mapping; to-date,
no claims for periodic phenomena on time-scales of tens of years have been made. However,
 T\"{u}rler \etal (2000) presented recently a long-term light curve of 3C273 in the
submillimeter to radio bands. Their Figure 3 shows these light curves from 0.45 mm 
down to 2.7 GHz and has two possible peaks more evident in the three bands from 37GHz to 15 GHz
that are spaced by 7-8 yrs, about half the photometric quasi-period suggested by Angione 
\& Smith (1985). T\"{u}rler \etal (2000) interpret the variability as the
signature of multiple outbursts in the synchrotron emission from the jet. If such
outbursts are periodic or quasi-periodic, the cause is likely to be a physical
mechanism such as  precession of the accretion disk and of the jets, as in SS433.

Additional optical variability work using more photographic plate data can be
applied to expand the work of Angione \& Smith (1985) so as to confirm or reject the
approximate period which they determined. In the
Wide-Field plate archive we found information on many hundreds of additional plates covering the
3C273 field that were not included in the Angione \& Smith work. The ready
availability of such data in digital form would clearly constitute an extremely
valuable resours for the determination of such basic scientific facts.

\begin{figure} 
\vspace{7cm}
\includegraphics{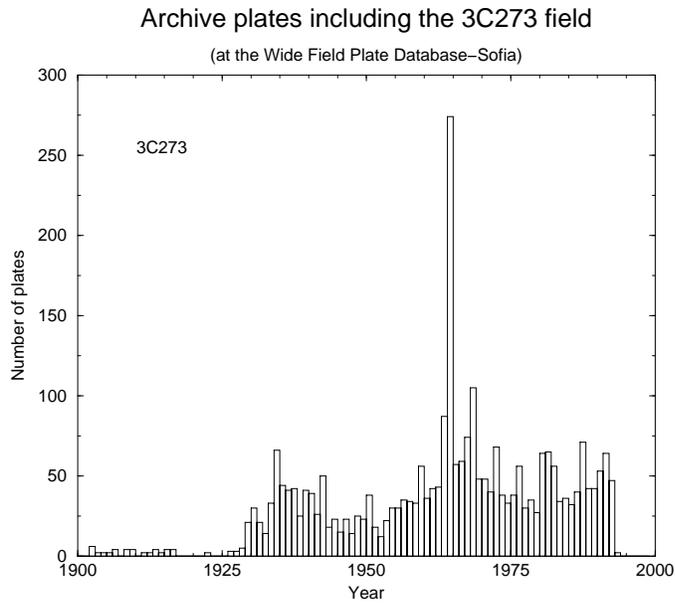}
\caption{Photographic plates covering the field of 3C273 available world-wide, as 
logged in the Wide-Field plate archive.}
\end{figure}

\section{Technical feasibility}
In order to realize the potential of the historical information recorded on astronomical
plates, it is vital to present that information in digital form for
inclusion in modern studies. To that end, the international community needs to:
\begin{enumerate}
\item locate historical plate archives and catalog their contents in a computer-readable format
\item provide appropriate storage conditions to safeguard the plates; this
may entail concentrating the plates at a few locations
\item ensure the accessibility of the information for research
\item select or design modes for scanning the relevant plates
\item generate reasonably complete digital datasets from the photographic information
\end{enumerate}

The first task is already well in hand, following the opening of the WFPA
in Sofia and the availability of computerized searches through the SIMBAD facility. 

An important step towards accomplishing the remaining tasks
was taken in March 2000, by a meeting at the Royal Observatory of Belgium (ROB), in 
Brussels. This institution has offered to host a plate archive for  European 
observatories, to be called UDAPAC (Uccle Direct Astronomical Plate Archive Centre).
The UDAPAC collection could ultimately contain 0.5 million plates, about one-quarter
of the entire world collection. To realize this part of the project, UDAPAC must
adapt and refurbish the assigned space offered by the ROB into
suitable plate vaults (waterproofing the walls, providing air conditioning, 
suitable storage cabinets, and the like).

The cheif objective of the project is the transfer of the photographic information
into digital information for use by the astronomical community. That requires the 
development and/or modification of scanners to make them (a) fast and (b) 
accurate enough for the various scientific tasks which need to be undertaken.
 At least two parallel approaches could be followed:
\begin{enumerate}

\item The High-Speed Scanner (HISS: Kroll \etal 1999) of Sonneberg Observatory could 
be copied or adapted, and made available to the 
ROB and to other similar facilities. The HISS  is a six-CCD scanner
that can digitize a sizeable plate (up to 30$\times$30 cm$^2$) 
in less than 15 minutes. First tests show the machine to be 
stable and accurate. The important point is that HISS was produced with a budget of less than
\$100k.

\item A commercial flatbed scanner could be adapted for astronomical work. High-quality scanners
are now available for less than \$50k. A set of such scanners was tested by the publishing industry
(Brunner \& Lindstr\"{o}m 1999) and the results are encouraging.UDAPAC is 
planning  tests of commercial scanners for astronomical plates.

\end{enumerate}

\section*{Acknowledgements}
We are grateful to Dr. Elizabeth Griffin, whose determined efforts brought about the 
adoption of IAU Resolution B3, supporting efforts to save astronomical plates. 
She and Alain Fresneau (the latter on behalf of the
IAU Working Group on 100-yr old plates) were instrumental in bringing about the Brussels
meeting where the European Plate Centre UDAPAC was started.

\section*{References}

\begin{description}

\item Angione, R.J., Roosen, R.G., Sievers, J. \& Moore, E.P. 1981, AJ, 86, 653

\item Angione, R.J. \& Smith, H.J. 1983, AJ, 90, 2474

\item Brunner, L. \& Lindstr\"{o}m. P. 1999, ``The Seybold Report on Publishing Systems'' {\bf
28}(9), 3 and {\bf 28}(11), 11

\item Bondar', N.I. 1995, A\&AS, 111, 259

\item Fresneau, A., Argyle, R.W., Marino, G. \& Messina, S. 2000, in press

\item Girard, T.M., Wu, H., Lee, J.T., Dyson, S.E., van Altena, W.F., Horch, E.P.,
Gilliland, R.L., Schaefer, K.G., Bond, H.E., Ftaclas, C., Brown, R.H., Toomey, D.W., 
Shipman, H.L., Provencal, J.L. \& Pourbaix, D. 2000, AJ, 119, 2428

\item Hollis, J.M., Bertram, R., Wagner, R.M. \& Lampland, C.D. 1999, ApJ, 514, 895

\item Hudec, R., Dedoch, A., Pravek, P. \& Borovicka, J. 1994, A\&A, 284, 839

\item Hudec, R. \& Wenzel, R. 1976, Bull. Astron. Soc. Czechoslovakia, 27, 325

\item Kroll, P., Brunzendorf, J. \& Fleischmann, F. 1999 in ``Treasure Hunting in Astronomical
Plate Archives'' (P. Kroll, ed.) Thun and Frankfurt am Main: Verlag Harri Deutsch, p.61

\item Pasachoff, J.M. \& Olson, R.J.M. 1995, AAS, 187, 3501

\item Schaeffer, B.E. 1990, ApJ, 364, 590

\item Standish, E.M. 1996, in ``Completing the inventory of the Solar System'' 
(T.W. Retting \& J.M. Hahn, eds.) Pasadena, CA.: Astronomical Society of the Pacific, p. 163

\item Simon, V., Hudec, R. \& Kroll, P. 2000, Proceedings of the 4th INTEGRAL workshop, 
Alicante, in press

\item Torres, G. \& Stefanik, R.P. 2000, ApJ, 119, 1914

\item Tsvetkov, M.K., Stavrev, K.Y., Tsvetkova, K.P., Semkov, E.H., Mutafov, A.S.
\& Michailov, M.-E. 1997, Baltic AStron., 6, 271

\item T\"{u}rler, M. Courvoisier, T.J.-L. \& Paltani, S. 2000, A\&A preprint,
astro/ph/0008480

\end{description}

\end{document}